\documentclass[conference]{IEEEtran}

\usepackage{graphicx}
\usepackage{subfigure}
\usepackage{amsmath}
\usepackage{braket}
\usepackage{algorithm, algorithmicx, algpseudocode}
\usepackage{comment}
\usepackage{tikz}
\usepackage{amssymb} % for checkmark symbol
\usepackage{booktabs} % for better horizontal rules
\usepackage{array} % for better vertical rules

\usepackage{listings}
\usepackage{seqsplit}
\usepackage{hyperref}
\usepackage{xcolor}

\usepackage{multirow}
\usepackage{pdfpages}

\usepackage{ulem}

\usepackage{babel}
\usepackage[T1]{fontenc}
\usepackage[utf8]{inputenc}

\newcommand{\todo}[1]{}
\renewcommand{\todo}[1]{{\color{red} TODO: {#1}}}

\begin{document}
% \linenumbers % Enable line numbering

\title{GPU-Accelerated Distributed QAOA on Large-scale HPC Ecosystems}

\author{
    % Anonymous Author(s)\\
    % For Double-Blind Review\\ \\    
    
    % commenting out before submission
   \IEEEauthorblockN{Zhihao Xu$^1$\IEEEauthorrefmark{1}\IEEEauthorrefmark{2}, Srikar Chundury\IEEEauthorrefmark{3}, Seongmin Kim$^2$\IEEEauthorrefmark{1}, Amir Shehata\IEEEauthorrefmark{1}, Xinyi Li\IEEEauthorrefmark{4}, \\ Ang Li\IEEEauthorrefmark{4}, Tengfei Luo\IEEEauthorrefmark{2}, Frank Mueller\IEEEauthorrefmark{3}, and In-Saeng Suh\IEEEauthorrefmark{1}} 
   \\
   \IEEEauthorblockA{\{$^1$xuz2; $^2$kims\}@ornl.gov} 
   \IEEEauthorblockA{\IEEEauthorrefmark{1}Oak Ridge National Laboratory, Oak Ridge, TN 37831, USA}
   \IEEEauthorblockA{\IEEEauthorrefmark{2}University of Notre Dame, Notre Dame, IN 46556, USA}
   \IEEEauthorblockA{\IEEEauthorrefmark{3}North Carolina State University, Raleigh, NC 27695, USA}
   \IEEEauthorblockA{\IEEEauthorrefmark{4}Pacific Northwest National Laboratory, Richland, WA 99354, USA}
  
}
 
% \author{
%   \IEEEauthorblockN{Anonymous Author(s)}
%   \IEEEauthorblockA{For Double-Blind Review}
% }

\maketitle

\begin{abstract}

Quantum computing holds great potential to accelerate the process of solving complex combinatorial optimization problems. The Distributed Quantum Approximate Optimization Algorithm (DQAOA) addresses high-dimensional, dense problems using current quantum computing techniques and high-performance computing (HPC) systems. In this work, we improve the scalability and efficiency of DQAOA through advanced problem decomposition and parallel execution using message passing on the Frontier CPU/GPU supercomputer. Our approach ensures efficient quantum-classical workload management by distributing large problem instances across classical and quantum resources. Experimental results demonstrate that enhanced decomposition strategies and GPU-accelerated quantum simulations significantly improve DQAOA’s performance, achieving up to 10$\times$ speedup over CPU-based simulations. This advancement enables better scalability for large problem instances, supporting the practical deployment of GPU systems for hybrid quantum-classical applications. We also highlight ongoing integration efforts using the Quantum Framework (QFw) to support future HPC-quantum computing systems.

\end{abstract}


\section{Introduction}

The Quantum Approximate Optimization Algorithm
(QAOA)~\cite{farhi2014quantumapproximateoptimizationalgorithm,
  Blekos_2024, kim2024performance} and its distributed variant,
DQAOA~\cite{kim2024distributed}, are among the most promising
quantum-classical hybrid algorithms to solve combinatorial
optimization problems exploiting contemporary noisy quantum
devices. However, executing these algorithms on near-term quantum
hardware remains challenging due to qubit decoherence, gate errors,
and limited qubit connectivity. As a result, high-fidelity quantum
circuit simulators play a crucial role in validating and optimizing
these algorithms before real-world deployment.

Quantum simulators such as Qiskit Aer~\cite{qiskit2024},
NWQ-Sim~\cite{nwqsim2025} (for full-state vector~\cite{9910084} and
density matrix simulations~\cite{9355323}), and
TN-QVM~\cite{10.1145/3547334} (for tensor network-based simulations)
allow researchers to model quantum circuits classically with high
precision. These simulators are essential for studying algorithmic
behavior, analyzing noise effects, and tuning quantum-classical hybrid
optimization workflows. However, they face significant computational
challenges due to the exponential growth in quantum state
representations. Full-state vector simulators, such as NWQ-Sim and
Qiskit Aer, require an exponential increase in memory and
computational complexity as the number of qubits increases, making
them impractical for large-scale simulations on conventional computing
systems~\cite{kim2025distributed}. Similarly, density matrix
simulations, which are crucial for modeling noise effects in quantum
circuits, demand even greater memory resources. Tensor network-based
simulators, such as TN-QVM, reduce the memory footprint by leveraging
the circuit structure and low entanglement, but tensor contractions
remain computationally expensive, especially for deep circuits.

Given these limitations, large-scale HPC systems are essential for
scaling quantum simulations beyond small problem sizes. HPC platforms
provide the necessary computational resources, including high memory
capacity, fast interconnects, and distributed computing capabilities,
to support large-scale quantum simulations. The ability to distribute
workloads across multiple nodes allows for efficient parallelization
of quantum state evolution and tensor contractions, making it possible
to simulate larger qubit systems. For example, high-speed interconnects, such as HPE
Slingshot~\cite{hpe-slingshot} and InfiniBand~\cite{infiniband},
enable fast communication between nodes, which is crucial for
distributed quantum simulations that require frequent data exchanges.

To further enhance the efficiency of quantum simulations,
GPU acceleration is
critical~\cite{cicero2024simulationquantumcomputersreview,
  WILLSCH2022108411}. Modern quantum circuit simulations rely on
computationally intensive linear algebra operations, including
matrix-vector multiplications, Kronecker products, and tensor
contractions, all of which are highly parallelizable. GPUs, such as
NVIDIA A100/H100 and AMD MI250X, offer massive parallelism, enabling
thousands of simultaneous computations that significantly accelerate
quantum state evolution~\cite{vallero2025statepracticeevaluatinggpu,
  10.1145/3624062.3624216, 10821080}. The high memory bandwidth of
GPUs, particularly those with HBM2e memory~\cite{hbm-gpu-memory},
allows rapid access and manipulation of quantum state vectors, which
can contain billions of complex numbers in large simulations. These
capabilities make GPUs an indispensable component for executing
large-scale quantum simulations efficiently.

Multi-GPU and multi-node scalability further enhance the feasibility
of distributed quantum simulations, particularly for
DQAOA~\cite{kim2024distributed}. Simulators like NWQ-Sim and TN-QVM
leverage multi-GPU execution to distribute workloads across multiple
accelerators within a single node or across multiple nodes using the
message passing interface (MPI)~\cite{10.1145/3547334,
  suh2024simulatingquantumsystemsnwqsim,
  li2024tanqsimtensorcoreacceleratednoisy}. This parallelization is
particularly beneficial for distributed QAOA, where large optimization
problems can be partitioned and processed concurrently across multiple
GPUs, aligning well with large-scale HPC architectures. The ability to
distribute computations efficiently ensures that hybrid quantum-HPC
workflows can take full advantage of modern supercomputing resources.

QAOA is a leading candidate for the near-term quantum
advantage~\cite{doi:10.1126/sciadv.adm6761}, providing a
quantum-classical variational framework for solving combinatorial
optimization problems. DQAOA extends this approach by distributing
problem instances across multiple quantum processors or simulators,
making it well-suited for integration with large-scale HPC
architectures. To evaluate the performance of DQAOA at scale, in this
work, we conduct benchmark tests comparing CPU-only and
GPU-accelerated simulations on the Frontier supercomputer. Our results
show that using GPUs provides an order of magnitude speedup (nearly 10$\times$) compared to running on CPUs. This performance gain is attributed to the superior ability of
GPUs to handle highly parallelizable linear algebra operations in
quantum circuit simulation, as well as their high memory bandwidth,
which significantly reduces data movement overhead.

These findings highlight the crucial role of GPU acceleration in
large-scale quantum-classical hybrid workloads and establish an
important benchmark for DQAOA implementations on exascale HPC
systems. Furthermore, we present single-node and multi-node
GPU-accelerated QAOA and DQAOA simulations, demonstrating scalability
and performance benchmarks on the Frontier supercomputer. Our
benchmark tests leverage CPUs and GPUs to evaluate computational
efficiency and multi-node performance for quantum
simulations. Additionally, to facilitate the seamless execution of
these hybrid workloads, one can leverage a Quantum Framework
(QFw)~\cite{shehata2024frameworkintegratingquantumsimulation} that
integrates with Qiskit and NWQ-Sim. This framework will optimize
workload distribution, enable efficient execution on multi-node GPU
systems, and provide a robust infrastructure for large-scale quantum
algorithm research.

%Need to summarize contributions at end of intro:
In summary, the contributions of this work are:
(1) The scalability and efficiency of DQAOA is improved via
large-scale HPC support with up to 160 CPUs/GPUs.
(2) GPU experiments indicate a speed of up to 10$\times$ over their
CPU counterpart.
(3) A quantum software framework is designed that provides distributed QAOA co-simulation with significant scalability capabilities.

\section{GPU-Accelerated Quantum Simulators}

\begin{figure}[!ht]
\centering
\includegraphics[width=1.0\linewidth]{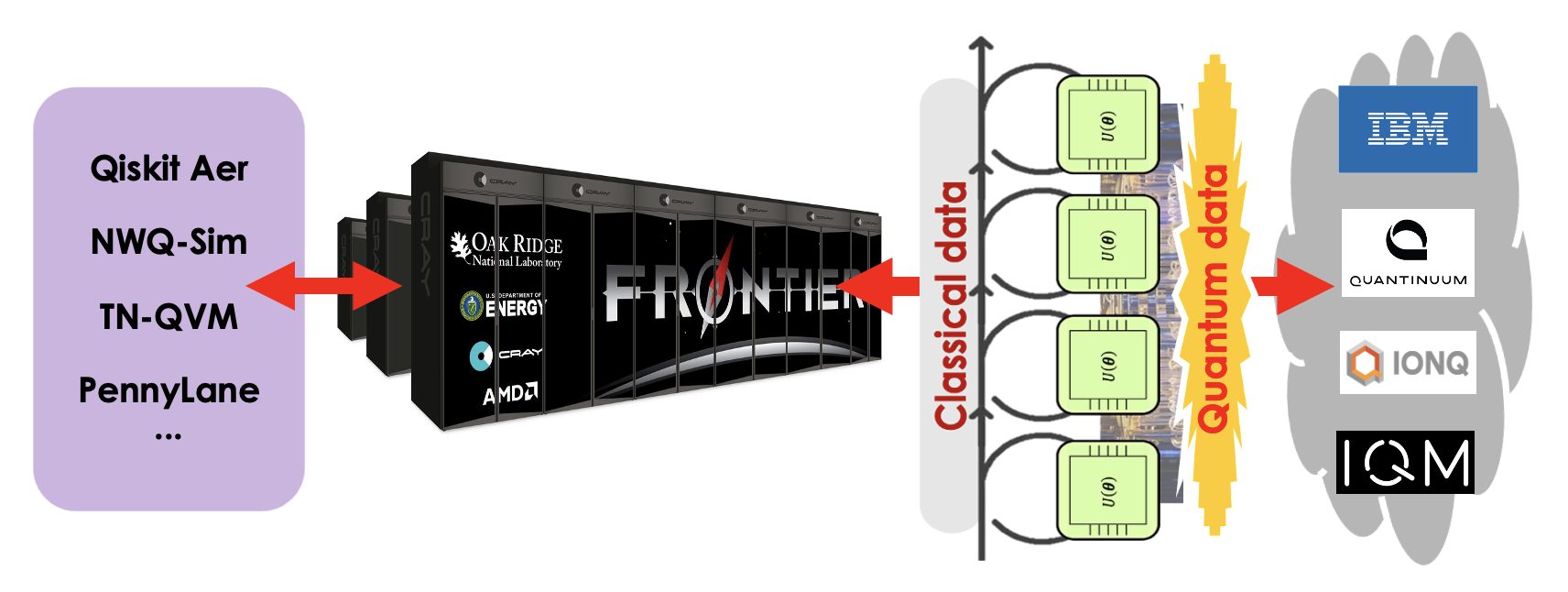}
\caption{\label{fig:frontier-qcup} Diagram of the Frontier
  supercomputer that integrates with quantum simulators and QCUP
  quantum devices.}
\end{figure}

The Frontier supercomputer~\cite{frontier}, located at Oak Ridge
National Laboratory, the world's first exascale system, serves as a
groundbreaking platform for high-performance computing (HPC), offering
unprecedented capabilities for large-scale computational applications,
including quantum simulation. Built on HPE Cray EX technology and
powered by AMD Instinct MI250X GPUs, Frontier delivers exceptional
processing power tailored for simulating quantum systems. Each compute
node is equipped with four MI250X GPUs, interconnected via AMD
Infinity Fabric, providing 200 GB/s of bidirectional bandwidth. With
64 GB of high-bandwidth memory (HBM2) per GPU and a peak
double-precision performance of 47.9 petaflops per GPU, Frontier
enables large-scale quantum simulations that push the boundaries of
computational science.  Additionally, Frontier is seamlessly
integrated with real quantum hardware through the Quantum Computing
User Program (QCUP)~\cite{qcup}, enabling access to state-of-the-art
quantum computing platforms such as IBM Quantum, Quantinuum, IonQ, and
IQM. Figure~\ref{fig:frontier-qcup} shows the diagram of the
large-scale quantum HPC ecosystem that integrates with quantum
simulators and QCUP quantum devices.

\subsection{Quantum Simulators on Frontier}

Frontier is optimized for a diverse range of quantum simulation
frameworks, including both quantum circuit-based and Hamiltonian-based
approaches. The integration with Qiskit Aer, NWQ-Sim, and TN-QVM allows
researchers to explore quantum algorithms, validate quantum hardware,
and benchmark emerging quantum computing technologies.

Qiskit Aer~\cite{qiskit2024} is a high-performance quantum circuit
simulator designed to support QASM, state vector, density matrix, and
extended stabilizer simulations. It takes advantage of Frontier's
GPU-accelerated dense linear algebra operations, allowing for
efficient simulation of quantum circuits. Using high-performance
tensor operations, Qiskit Aer enables researchers to model quantum
circuits with realistic noise characteristics, making it a valuable
tool for assessing the performance of quantum error correction
protocols and NISQ (Noisy Intermediate-Scale Quantum) devices.

\begin{figure}
    \centering
    \includegraphics[width=0.95\linewidth]{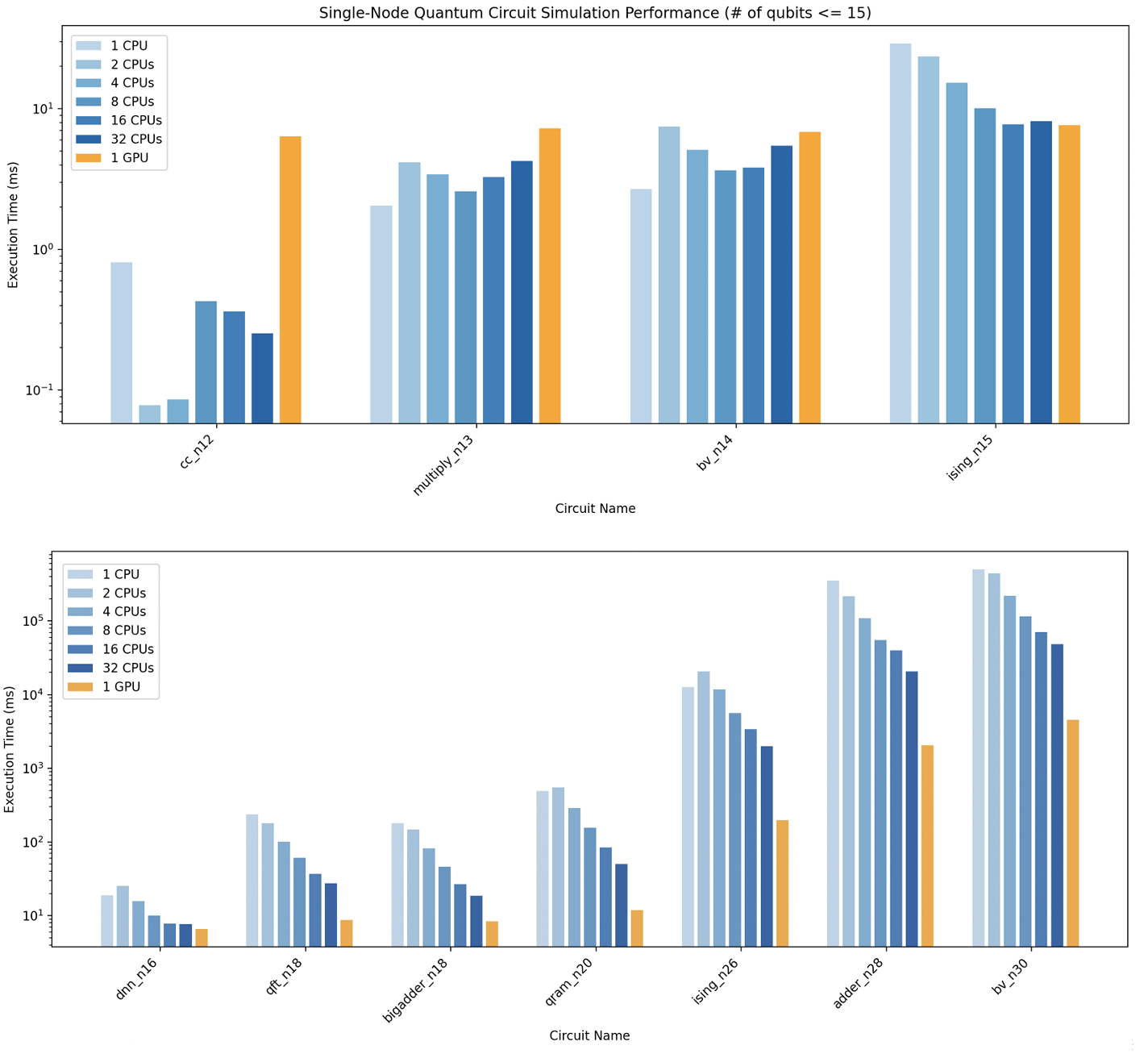}
    \caption{QASMBench~\cite{10.1145/3550488} with NWQ-Sim on Frontier's CPUs and GPU.}
    \label{fig:qasmbench-gpu}
\end{figure}

% NWQ-Sim~\cite{nwqsim2025} is specifically designed for state
% vector~\cite{9910084} and density matrix simulations~\cite{9355323},
% optimizing Frontier's GPU-accelerated numerical linear algebra
% routines to model large-scale quantum systems, including those
% incorporating quantum noise effects. These capabilities are crucial
% for validating quantum algorithms and developing noise-resilient
% quantum computing techniques. Figure~\ref{fig:qasmbench-gpu} shows
% benchmark tests for QASMBench~\cite{10.1145/3550488} with NWQ-Sim on
% Frontier's CPUs and GPU.
NWQ-Sim~\cite{nwqsim2025} is specifically designed for state vector~\cite{9910084} and density matrix simulations~\cite{9355323}, optimizing Frontier's GPU-accelerated numerical linear algebra routines and utilizing matrix accelerators to efficiently model large-scale quantum systems, including those with quantum noise effects. These capabilities are essential for validating quantum algorithms and developing noise-resilient quantum computing methods. Figure~\ref{fig:qasmbench-gpu} presents benchmark results using QASMBench~\cite{10.1145/3550488}, highlighting NWQ-Sim's superior performance on Frontier's GPUs compared to multi-core CPUs, especially as quantum circuit sizes increase, where even a single GPU significantly outperforms a CPU-based implementation.

TN-QVM (Tensor Network Quantum Virtual Machine)~\cite{10.1145/3547334}
is optimized for tensor network-based quantum simulations, efficiently
utilizing GPU-accelerated tensor contractions to model high-depth
quantum circuits with significant entanglement. This makes it
particularly effective for simulating many-body quantum systems and
investigating complex quantum phenomena, such as quantum phase
transitions and topological states of matter.

These advanced quantum simulation frameworks play a critical role in
the ongoing development and validation of quantum algorithms. They
serve as essential tools for benchmarking near-term quantum
processors, enabling researchers to compare simulated results with
real quantum hardware performance (see \cite{kim2024distributed}). 
By complementing physical quantum
processors with large-scale simulation capabilities, Frontier is
helping to bridge the gap between current noisy intermediate-scale
quantum (NISQ) devices~\cite{Preskill2018quantumcomputingin} and the
development of fault-tolerant quantum computing in the future.

Frontier's exascale computational power is used for pre-processing,
error mitigation, and classical optimization, all of which are crucial
to improving the efficiency of quantum algorithms. For example, Qiskit
Aer's noise-aware simulation \cite{suh2024simulatingquantumsystemsnwqsim} 
features allow researchers to fine-tune
quantum circuits before executing them on actual quantum hardware,
significantly enhancing their robustness against real-world
decoherence and gate errors.

In addition, Frontier's support for mixed-precision computing provides
a key advantage in large-scale quantum simulations \cite{li2024tanqsimtensorcoreacceleratednoisy}. 
By dynamically adjusting precision levels, researchers can accelerate tensor
operations while ensuring numerical accuracy where needed, leading to
more efficient computations across a broad spectrum of quantum
applications.

\subsection{Hybrid Quantum-HPC Workflows}

The integration of Frontier and QCUP devices enables the development
of hybrid quantum-HPC workflows that leverage the complementary
strengths of classical and quantum computing, addressing complex
computational challenges across diverse scientific
domains~\cite{ALEXEEV2024666, BECK202411}. By combining the
scalability and precision of high-performance classical computing with
the potential advantages of quantum algorithms, hybrid
quantum-classical approaches offer a promising pathway to solving
problems that are intractable with classical methods alone.  This
synergy is crucial for advancing research in fields such as quantum
simulations, optimization, and materials science, where hybrid
methodologies can significantly enhance computational efficiency and
solution accuracy.

Hybrid quantum-classical computing is essential for tackling
computational challenges across various disciplines:
\begin{itemize}
\item Quantum
  chemistry~\cite{robledomoreno2024chemistryexactsolutionsquantumcentric}
  --- Simulating molecular structures and reaction mechanisms with
  greater accuracy than classical models.
\item Materials discovery~\cite{ALEXEEV2024666,
    lourenco2024exploringquantumactivelearning} --- Exploring novel
  materials with unique quantum properties for energy storage,
  superconductivity, and catalysis.
\item Quantum machine
  learning~\cite{grana2025materialsdiscoveryquantumenhancedmachine}
  --- Developing hybrid algorithms that enhance classical AI models with
  quantum advantages.
\item Optimization~\cite{kim2024distributed, Abbas_2024} --- Enhancing
  classical optimization techniques with quantum approaches to solve
  complex combinatorial problems, such as logistics, scheduling,
  financial modeling, and large-scale scientific simulations, more
  efficiently.
\end{itemize}
Thus, by bridging quantum and classical computing, hybrid workflows enable
more efficient problem-solving strategies, unlocking new possibilities
for scientific discovery and technological innovation.

In this work, we utilize GPU-accelerated simulators, Qiskit Aer and NWQ-Sim, to evaluate hybrid workflows for DQAOA on the Frontier CPU/GPU systems. Future efforts will include the implementation of DQAOA on the GPU-based TN-QVM simulator. The workloads on QCUP quantum devices are beyond the current scope of this study; related artifacts are available in~\cite{kim2024distributed}.

\section{A Quantum Framework for Large-scale HPC Ecosystems}

% \textcolor{blue}{Amir, Srikar}
\begin{figure}[!htb]
   \centering
   \includegraphics[width=0.85\columnwidth]{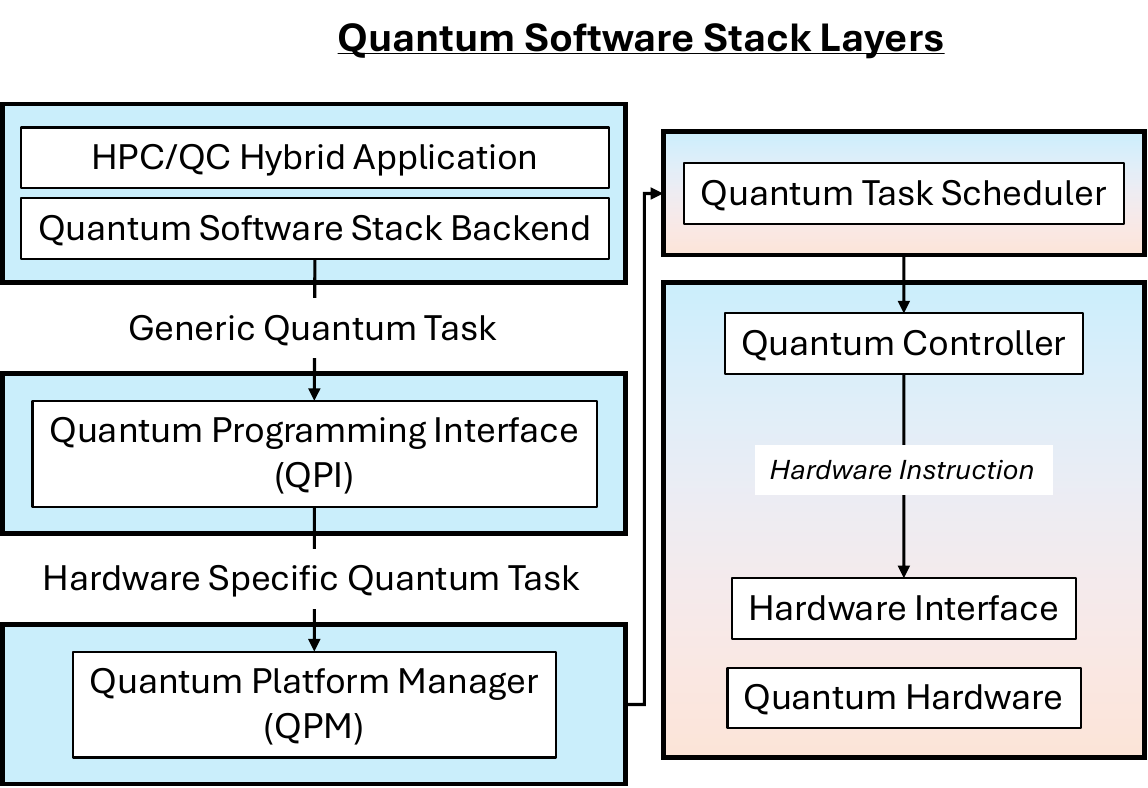}
   \caption{Enhanced quantum software stack depicting
     abstraction layers that separate the application interface from
     low-level tools, some of which are hardware specific. }
   \label{fig:qstack}
\end{figure}

The Quantum Framework
(QFw)~\cite{shehata2024frameworkintegratingquantumsimulation,
  shehata2025buildingsoftwarestackquantumhpc} is a prototype that
implements the software stack, depicted in Figure~\ref{fig:qstack},
which is designed to abstract low-level platform-specific operations
from applications. It achieves this by introducing multiple software
layers, each providing an appropriate level of abstraction to the
layer above it. At the highest level, the HPC/QC hybrid application
forms the topmost layer. The application executes on classical HPC
resources after the requested resources have been allocated. Resource
allocation is managed by a quantum-aware resource manager, ensuring
efficient distribution of both classical and quantum computing
resources.

There are two supported allocation methods: simultaneous and
interleaved~\cite{shehata2025buildingsoftwarestackquantumhpc}. In
simultaneous allocation mode, both quantum and classical resources are
reserved concurrently. This is ideal for applications that require
executing multiple small circuits in parallel and using the results
for classical post-processing.  The interleaved allocation mode is
best suited for workloads with a strict separation between quantum and
classical steps. For example, applications can pre-process data,
prepare quantum circuits for execution, run these circuits, and then
process the results. In this mode, quantum and classical steps are
interleaved such that only the required resources, either classical or
quantum, are reserved at any given time. QFw supports both
allocation modes.

Many quantum applications use existing circuit-oriented software
packages, such as Qiskit, to construct and run quantum circuits. To
support these, QFw introduces the ``Quantum Software Stack
Backend'' layer, which is designed to target such packages. This
allows applications to be deployed in an HPC environment with minimal
modifications; users simply switch to the QFw backend instead of their
current backend.

The QFw backend interacts with the Quantum Programming Interface (QPI)
to query and manage quantum resources, execute circuits, and handle
results and errors. The QPI provides a set of APIs for these tasks,
enabling a generic quantum circuit to be converted into an
intermediate representation such as QASM or QIR. Circuit
transformation and optimization tools, e.g., to reduce the number of
gates, can then be applied before transpiling and converting the
circuit to a hardware-specific format. The final circuit is passed
down to the hardware for execution via the Quantum Platform Manager
(QPM), which offers standardized APIs for hardware interaction. Each
hardware provider can implement the QPM APIs according to their
specific requirements.

\begin{figure}
    \centering
    \includegraphics[width=1.0\linewidth]{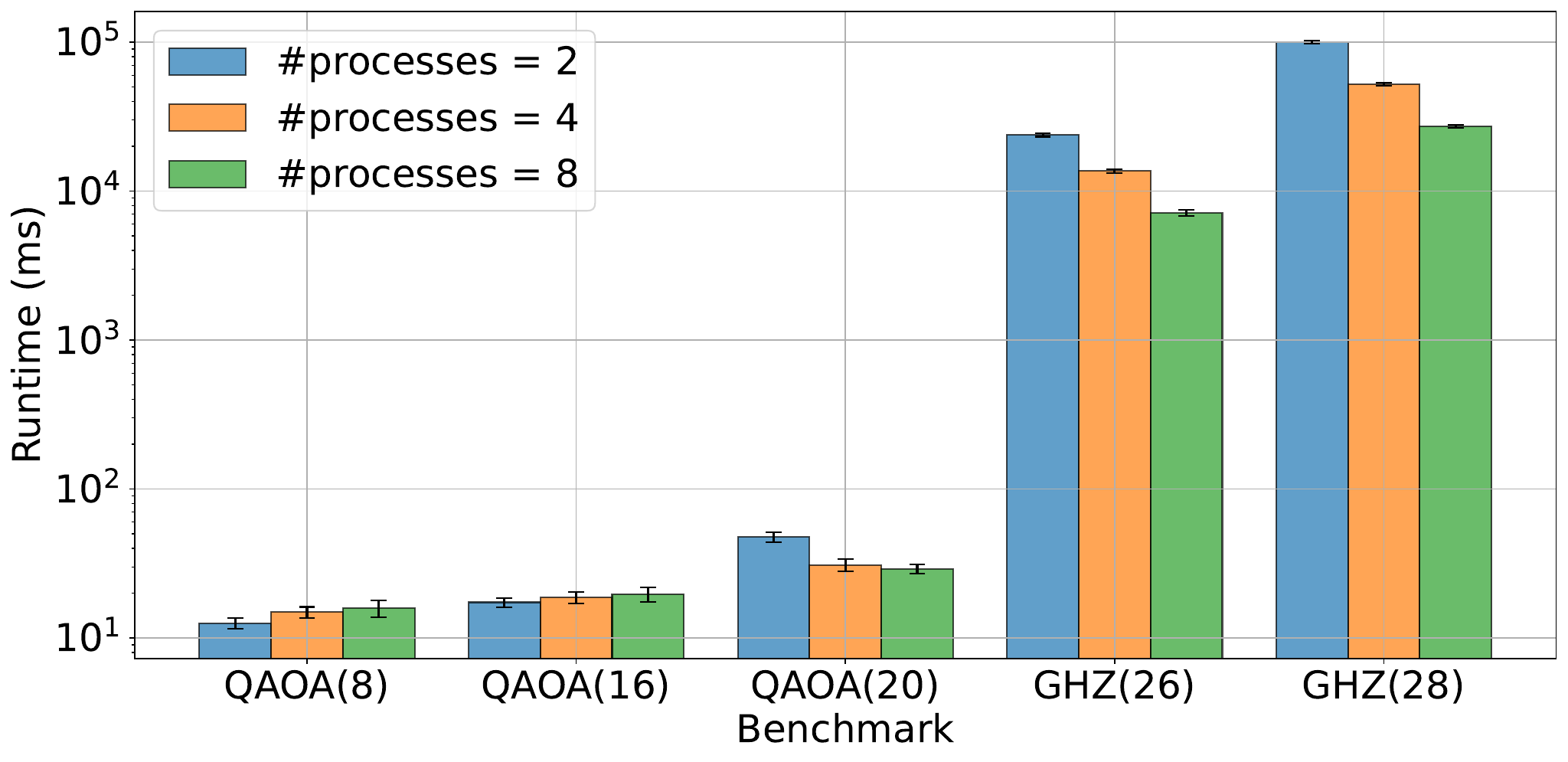}
    \caption[Single-node simulations with QFw]{Single-node QFw runs using NWQ-Sim for GHZ and QAOA.}
    \label{fig:single-qfw}
\end{figure}

QFw also supports running quantum circuits on classical quantum
simulators. It allocates two sets of HPC nodes, one for the classical
application and another for the quantum simulators. Currently, it
supports the TN-QVM, NWQ-Sim, and Qiskit-Aer simulators. Applications
can submit one or more circuits for execution, which are then
distributed across the HPC nodes dedicated to the quantum
simulators. If the selected simulator supports MPI, large circuit
simulations can span multiple nodes to leverage distributed computing.
As an illustration, Figure~\ref{fig:single-qfw} highlights how QFw enables straightforward and scalable execution of a variational algorithm (QAOA) and a single-run algorithm (GHZ) using NWQ-Sim on a single CPU node. This demonstrates QFw’s flexibility in managing different workloads with varying MPI parallelization levels.

% %
% \todo{The results indicate good scalability in terms of runtime as the
%   number of MPI tasks increases.}
% %
% \fm{How good? Linear with the number of tasks, with a 2$times$
%   speedup, or how much precisely?}
% %
% \fm{Can you produce results with more MPI processes, over multiple
%   nodes? This would better match the distributed QAOA claim and HPC simulation.} 
% %  
% \sri{I think our intention is to just show flexibility here, scalability is discussed in the QFw's section separately.}
% %

QFw simplifies the implementation of hybrid HPC/QC applications by
enabling quantum execution offloading with minimal application
changes. It allows users to seamlessly switch between different
simulators and real quantum hardware while providing a unified
platform for integrating simulators, hardware, and quantum tools.

% ORNL is actively collaborating with the community to establish a set of standard interfaces that will facilitate interoperability between different layers of the quantum software stack. Additionally, ORNL is focused on optimizing the software stack and exploring resource management strategies to ensure the efficient utilization of quantum and classical resources.

% \begin{figure}[!h]
% \centering
% \includegraphics[width=0.7\linewidth]{Figures/ghz_singlenode_qfw.png}
% \caption{\label{fig:single-ghz-qfw} Simulations of the GHZ quantum circuit on a single CPU and GPU processor with the GHZ and QFw integrated framework.}
%  \label{fig:ghz-qfw}
% \end{figure}
% \begin{figure}[!h]
% \centering
% \includegraphics[width=0.7\linewidth]{Figures/qaoa_singlenode_qfw.png}
% \caption{\label{fig:single-qaoa-qfw} Simulations of the QAOA on a single CPU and GPU processor with the QAOA and QFw integrated framework.}
%  \label{fig:qaoa-qfw}
% \end{figure}

In this work, we demonstrate the integration of DQAOA with QFw and present the design of its execution workflow. Although the full integration with DQAOA is still in progress, we provide its feasibility and showcase the initial integration of QAOA and the GHZ state simulations with QFw.

\section{\textbf{QAOA for Optimization Problems}}

Combinatorial optimization problems can be mapped to quadratic
unconstrained binary optimization (QUBO)
problems~\cite{kim2024review}. Quantum computing has been widely
employed to solve real-world optimization problems represented as
QUBOs~\cite{kim2022high, kim2024wide}. In particular, QAOA has shown
great promise in accelerating the solving process of QUBO
problems~\cite{doi:10.1126/sciadv.adm6761}, which
leverages parameterized quantum state \( |\psi(\theta)\rangle \),
where \( \theta \in \mathbb{R}^d \) represents the set of variational
parameters. The optimization process involves minimizing a cost
function \( \mathcal{C}(\theta) \) using a classical optimization:

\begin{equation}
\min_{\theta \in \mathbb{R}^d} \mathcal{C}(\theta).
\end{equation}  

The quantum state \( |\psi(\theta^*)\rangle \) is prepared based on
the determined optimal parameters \( \theta^* \) through a classical
optimization process. The bit string corresponding to the quantum
state is obtained after measurement, which can be used as a candidate
solution. Here, the QAOA ansatz is given by:
\begin{equation}
|\psi(\theta)\rangle = \prod_{j=1}^{p} e^{-i H_X \beta_j} e^{-i H \gamma_j} |+\rangle,
\end{equation}  
where \( p \) defines the circuit depth (number of layers), and
\( \beta_j, \gamma_j\) are real-valued variational parameters. In this
equation, \( H \) represents the problem Hamiltonian, and
$H_X = - \sum_{i=1}^{n} X_i$ is as a mixing Hamiltonian, with
\( X_i \) denoting the Pauli-\( X \) operator.  By mapping a QUBO
problem to the problem Hamiltonian (\textit{H}), QAOA can be used to
identify a bit string that minimizes the QUBO energy, making it
well-suited for minimization optimization problems.

We utilize Qiskit (v1.4.2), Qiskit-optimization (v0.6.1), and Qiskit-algorithms (v0.3.1) to construct the
QAOA ansatz. The Qiskit-aer (v0.15.1) and NWQ-Sim (v2.0) simulators serve as the backend to execute the QAOA circuits. To optimize
the variational parameters in the circuit, which are initialized with
random values, we employ COBYLA, a gradient-free optimizer. In this
study, the quantum circuit is implemented as a single layer (i.e.,
\textit{p} = 1).

\section{\textbf{Distributed QAOA for Solving Large-Scale Optimization Problems}}

As problem size increases, the number of pairwise interactions in QUBO
problems grows combinatorially, making it significantly more challenging
for QAOA to handle large-scale, dense problems because the two-qubit gate
operations expressing an interaction pair are computationally
expensive~\cite{kim2024performance}. This
scalability limitation restricts its practical applicability. To
address this issue, DQAOA has been introduced as a
scalable approach for solving large, dense QUBO problems
efficiently~\cite{kim2024distributed}.

\begin{figure}
    \centering
    \includegraphics[width=1\linewidth]{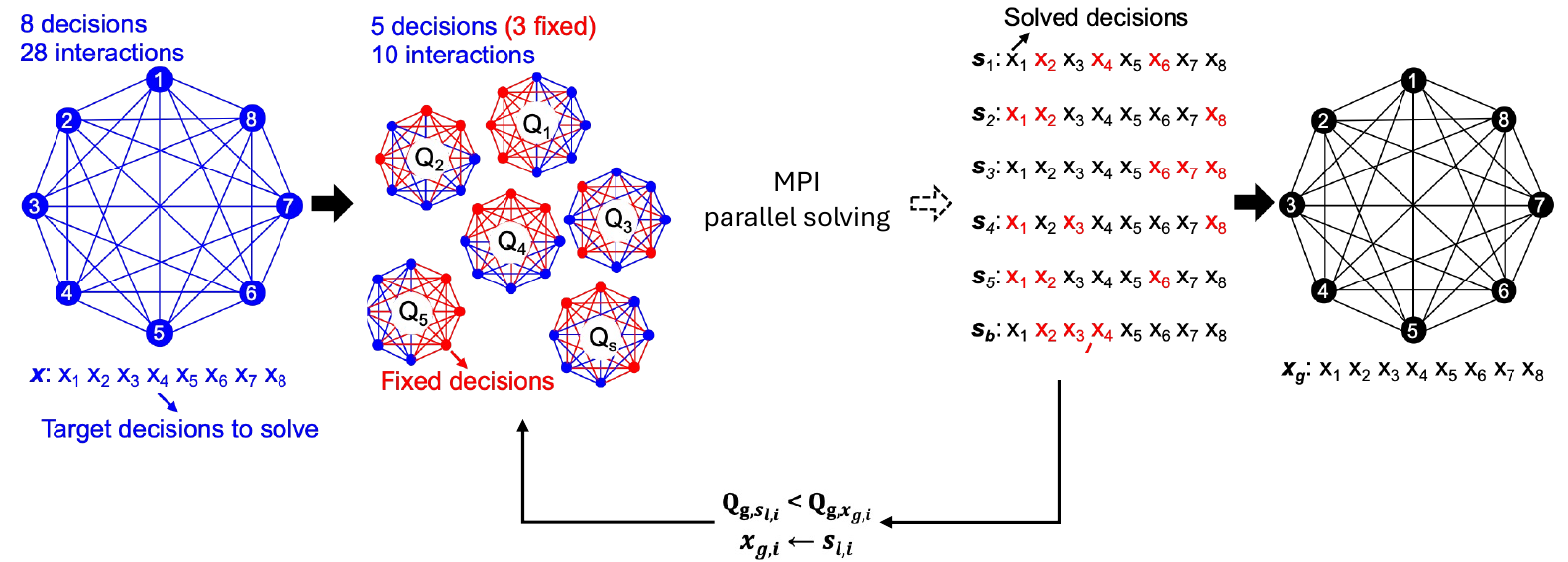}
    \caption[The schematic of DQAOA.]{Schematic of DQAOA to solve
      large-scale optimization problems through MPI distributed
      computing. A large QUBO is decomposed into $b$ sub-QUBOs, which
      are solved in parallel. The figure is adapted from
      ref.\cite{kim2024performance}.}
    \label{fig:dqaoa-mpi}
\end{figure}

Figure~\ref{fig:dqaoa-mpi} illustrates the workflow of DQAOA with
MPI. In DQAOA, large QUBO problems are decomposed into smaller
sub-problems, each of which is solved by QAOA on individual processors
within the HPC system. MPI serves as a communicator~\cite{RN241} that
efficiently sends and receives sub-QUBOs and sub-solutions through
distributed processors (CPUs and GPUs) asynchronously, so that we can
achieve high efficiency and parallel large-scale problem solving on
the HPC system.

As depicted in Figure~\ref{fig:dqaoa-mpi}, the DQAOA workflow includes three steps: 
\begin{enumerate}
\item \textbf{Problem decomposition:} A large QUBO problem is
  partitioned into smaller sub-QUBOs that match the capacity of QPU or
  classical processors. The decomposition methods are expected to
  capture the local structure of the original problem. This ensures
  that essential interactions from the original problem are preserved
  to achieve high approximation ratios within a few iterations. The
  decomposition method will be discussed in the next section.
\item \textbf{Parallel Processing with HPC:} Distributed parallel
  computation across multi-core and multi-node architectures on the
  Frontier enables the simultaneous solving of multiple sub-problems. This
  parallelization leads to significant speedup compared to a
  single-core setup.
\item \textbf{Aggregation of sub-solutions:} A global solution is
  iteratively updated by aggregating all sub-solutions from
  distributed processors. Specifically, fragments of the original
  solution will be updated by the corresponding sub-solution, and only
  those updates that lead to lower energy are accepted. The time
  complexity of the aggregation operation is $O(mkN^2)$, where $m$ is
  the number of sub-QUBOs, $k$ is the sub-QUBO size, and $n$ is the
  size of the original QUBO problem.
\end{enumerate}

\subsection{Decomposition Strategies}

As mentioned above, when solving a large-scale QUBO problem, one
strategy is to decompose the problem into sub-problems and leverage
parallel computing to exploit a divide-and-conquer strategy for the
original problem via DQAOA. However, the performance of DQAOA is
highly reliant on the decomposition quality~\cite{RN242,RN243}. For
example, if the decomposition disrupts significant interaction terms
in the QUBO matrix, it can introduce artificial dependencies,
redundant computations, or excessive aggregation costs. In addition,
if the decomposed sub-QUBOs include too many trivial decision
variables, the sub-solutions may fail to guide the global solution
toward the optimum. To maximize efficiency, decomposition should align
with the intrinsic local structures of the problem, minimizing
dependencies while enabling parallelism and effective problem-solving
at each recursive step.

Various decomposition methods exist for partitioning large QUBO
problems, which include random decomposition \cite{kim2024distributed}, and impact factor
directed (IFD) decomposition\cite{dwave_decomposition_2024}. These decomposition methods
significantly reduce quantum circuit depth and the number of two-qubit
gates, thereby improving the approximation ratio and reducing the
time-to-solution.

\subsubsection{Random Decomposition}

The random decomposition method decomposes a QUBO by selecting subsets
of nodes randomly based on the desired sub-QUBO size. Since this
method selects nodes randomly, it is robust against local minima but
may suffer from slower convergence due to the lack of structured
partitioning.

\subsubsection{Impact Factor Directed Decomposition}

One of the decomposition strategies proposed in this work is based on
the impact factor of each decision variable, which is called impact
factor directed (IFD) decomposition. The impact factor of a decision
variable is calculated by evaluating the change in the objective
function value when the variable is flipped in the current
solution. Essentially, the calculation of the impact factor is the
local sensitivity analysis for the current solution in the solution
space. Decision variables with higher impact factors are expected to
guide the optimization toward regions with steeper gradients, thereby
accelerating convergence. Consequently, these variables are given
higher priority for sampling and are included in sub-problems. The
details of IFD decomposition are given in Algorithm~\ref{alg:ifd}. The
complexity of the IFD decomposition is $O(mk^2)$, and the complexity
of the impact factor calculation is $O(N^3)$.

\begin{algorithm}[h!]
\caption{Impact factor directed (IFD) decomposition algorithm}\label{alg:ifd}
\begin{algorithmic}[1]
\Require $Q$ (QUBO matrix), sOpt (Current optimal solution), nSubQ (Number of sub-QUBOs), sizeSubQ (Size of each sub-QUBO)
\Ensure listSubQ (List of sub-QUBO matrices)
\State \textbf{Initialize:} listSubQ $\gets$ [ ]
\State indices $\gets$ \textbf{impact\_analysis}($Q$, sOpt) \Comment{Rank decision variables by impact factors}
\For{$i = 1$ to nSubQ}
    \State choice $\gets$ indices[i : i + sizeSubQ]
    \State \textbf{sort}(choice)
    \State subQ $\gets$ \textbf{zeros}(sizeSubQ, sizeSubQ)

    \For{$sub\_i = 1$ to subQUBO\_size}
        \For{$sub\_j = sub\_i$ to subQUBO\_size}
            \State subQ[$sub\_i, sub\_j$] $\gets$ $Q$[choice[$sub\_i$], choice[$sub\_j$]]
        \EndFor
    \EndFor
    \State \textbf{append} subQ to listSubQ
\EndFor
\State \Return listSubQ
\end{algorithmic}
\end{algorithm}

\begin{figure}
    \centering
    \includegraphics[width=0.8\linewidth]{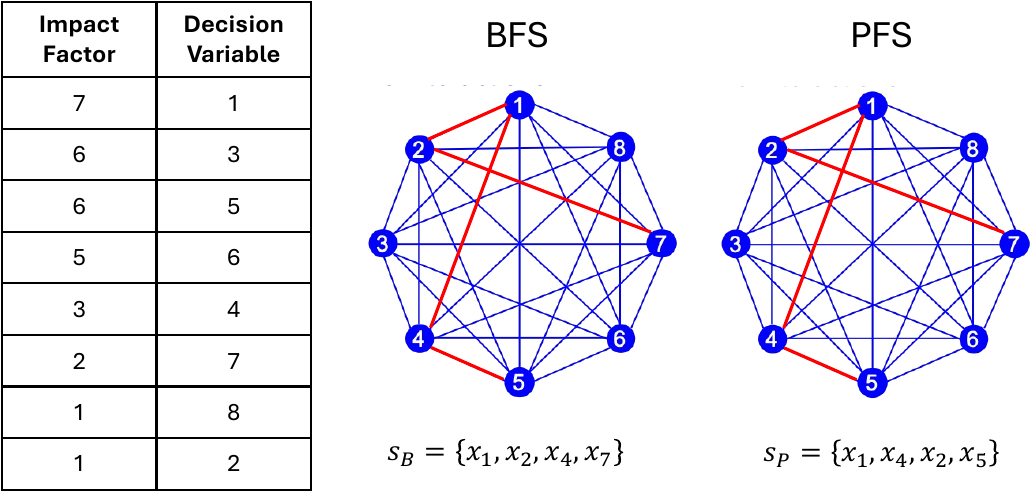}
    \caption[BFS and PFS for sub-QUBO extraction]{BFS and PFS for
      sub-QUBO extraction. The sample QUBO problem has 4 significant
      stronger interactions between decision variables ($x_1$, $x_2$),
      ($x_1$, $x_4$), ($x_2$, $x_7$) and ($x_4$, $x_5$), highlighted
      in red. $x_1$ is the start point for both BFS and PFS. BFS
      searches neighbor $x_2$ first, while PFS tends to travel toward
      $x_4$ first due to the higher impact factor.}
    \label{fig:bfs-pfs}
\end{figure}

Variables selected solely by impact factors may not be directly
connected in the sparse problems and might not represent local
structures. Therefore, this work also configures a mode of traversal
that not only considers the node impacts, but also captures features
that represent local structures within a problem. Breadth-first search
(BFS)~\cite{RN244} and priority-first search (PFS)~\cite{RN245} are two
graph traversal techniques that have been widely used in the community
of computer science, and play crucial roles in various applications,
including pathfinding~\cite{RN251}, network analysis~\cite{RN252}, and
artificial intelligence~\cite{RN253}. Considering the structural
similarity between QUBO and graphs, using BFS and PFS for large QUBO
decomposition can capture significant pair interactions between
decision variables~\cite{RN243}, especially in sparse QUBO
matrices. Figure~\ref{fig:bfs-pfs} shows an example of how BFS and PFS
work on extracting sub-QUBOs. BFS starts with the decision variable
with maximal impact factor, and then proceeds to strongly interacted
variables; while in PFS, the traversal selects the variables with the
highest impact factor among unselected decision variables that
strongly interacted with any already selected variable.

\section{\textbf{Results}}

\subsection{Comparison between different decomposition methods}

\begin{figure*}[!h]
    \centering
    \includegraphics[width=0.8\linewidth]{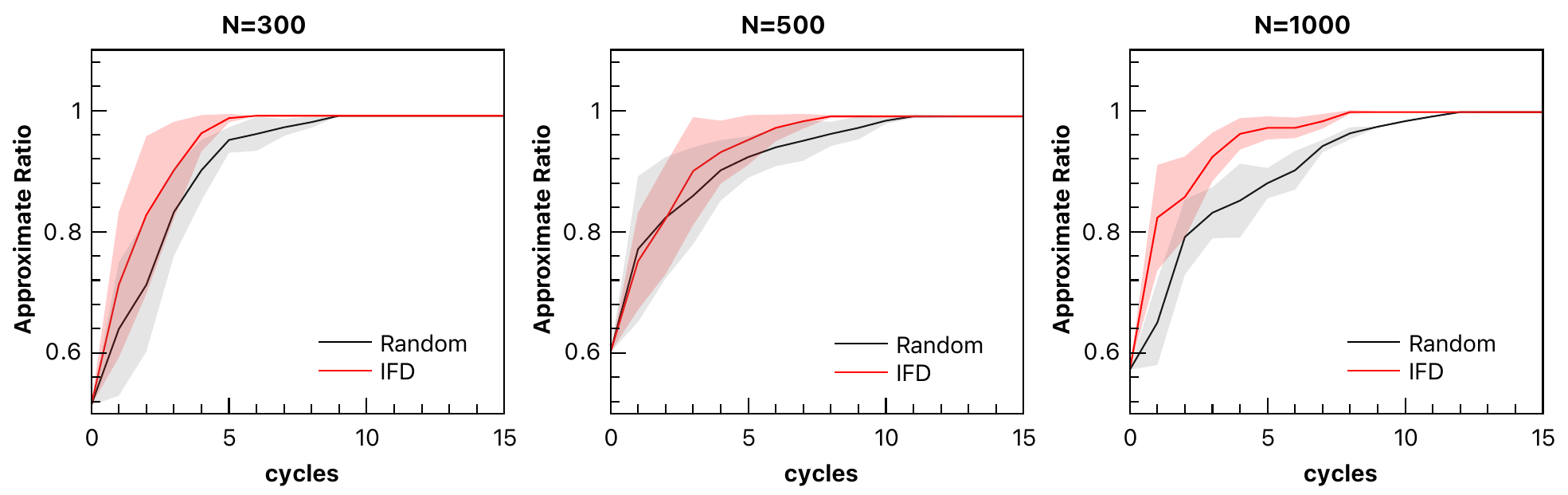}
    \caption{Comparison between different problem decomposition methods.}
    \label{fig:decompose}
\end{figure*}

Compared with the random sampling used in the previous DQAOA
study~\cite{kim2024distributed}, the IFD method can more effectively capture the decision variables that have significant impact on the global solution and apply limited computing resources to the most important sub-problems. Figure~\ref{fig:decompose} compares the performance of DQAOA using random and IFD decomposition methods for solving fully connected QUBO problems ranging from 300 to 1000 dimensions. All QUBO instances originate from metamaterial design problems. Approximate Ratio (A.R.) is used to evaluate the accuracy of DQAOA, which represents the ratio between energy from DQAOA and hybrid quantum annealing (HQA), which is regarded as the benchmark \cite{kim2025qabench}. All the results are collected from 10 independent trials with the same initialization. For the three cases, the sub-QUBO size and number of sub-QUBOs are set as 20 and 160, respectively, and DQAOA simulations are run with 160 CPU processors. As shown in Fig.~\ref{fig:decompose}, IFD always achieves faster convergence than random decomposition, especially when the QUBO problem is large.

Unlike fully connected QUBOs, sparse QUBO problems, such as Max-Cut
and the TSP, introduce distinct computational challenges. Although
sparse QUBO matrices contain fewer non-zero entries, their sparsity
does not necessarily imply lower computational challenge. Instead,
sparse QUBOs often exhibit highly fragmented energy landscapes, posing
significant difficulties for both classical and quantum algorithms. In
this context, effective problem decomposition becomes even more
critical to achieving computational efficiency and high-quality
solutions.

Max-Cut problems belong to a fundamental category of binary
combinatorial optimization problems \cite{kim2025qabench}. Given an undirected graph with
weighted or unweighted edges, the objective is to partition the set of
nodes into two groups such that the sum of the weights of the edges
between the two groups is maximized. As an NP-hard problem, Max-Cut is
computationally challenging to solve exactly, especially for large
graphs. Max-Cut problems can be formulated as QUBO problems, making them
suitable for quantum optimization techniques. As the number of nodes
(\textit{n}) increases, the possible number of edges grows
significantly, reaching a maximum of \textit{n}(\textit{n}-1)/2. The
problem's complexity (i.e., sparsity) can be controlled by adjusting
the sparsity of Max-Cut instances, which depends on the number of
edges. In this study, we set the number of edges to
\textit{n}(\textit{n}-1)/8, meaning the resulting QUBOs exhibit high
sparsity (25\%) compared to fully connected QUBOs.

\begin{figure}[!h]
    \centering
    \includegraphics[width=0.98\linewidth]{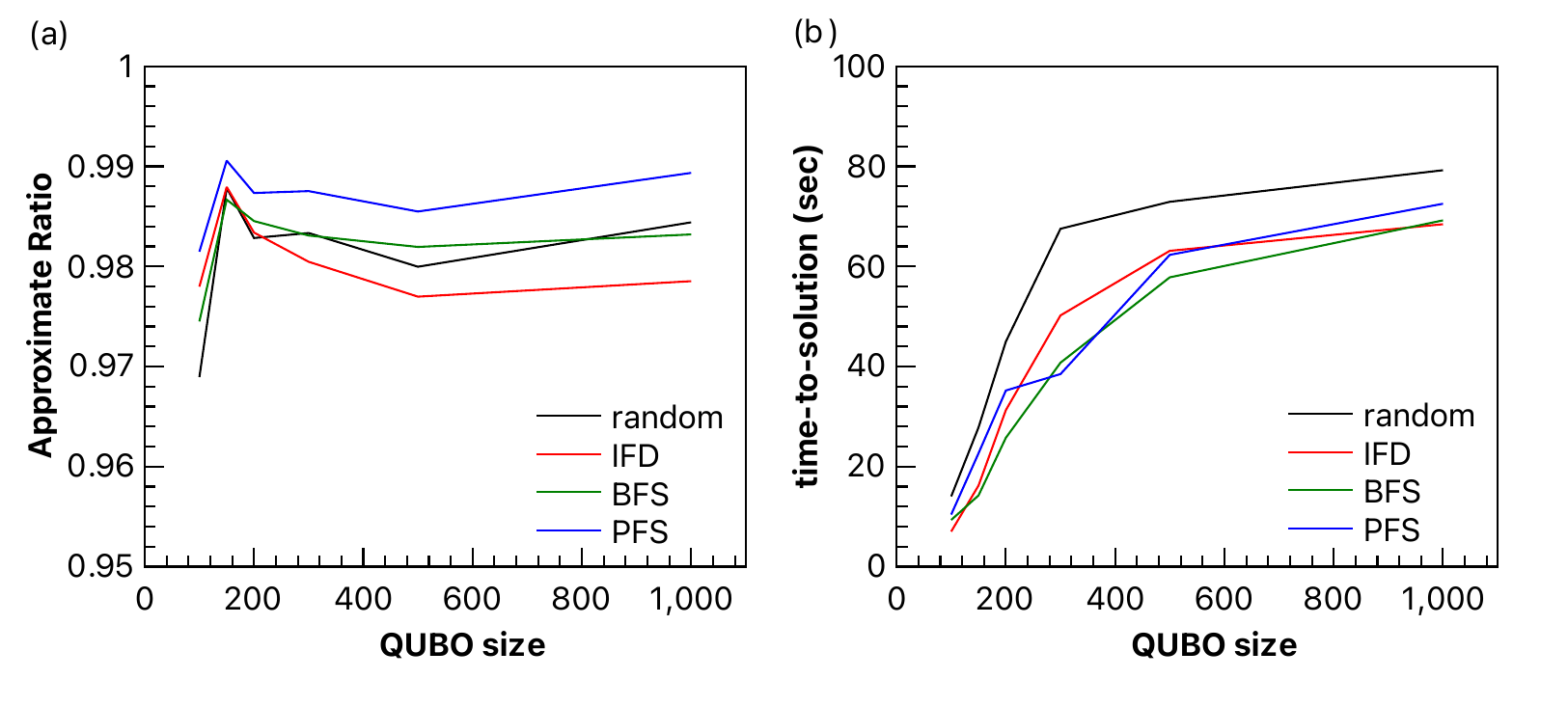}
    \vspace*{-\baselineskip}
    \caption{Comparison of DQAOA performance on Max-Cut problems using
      different decomposition methods: random, IFD, BFS and PFS. (a)
      Approximate ratio; (b) Time-to-solution.}
    \label{fig:mxc}
\end{figure}

Figure~\ref{fig:mxc} compares the performance of DQAOA when using different problem decomposition methods to solve the Max-Cut problem. We use random, IFD, BFS and PFS decomposition for DQAOA to solve the Max-Cut problem from 100 to 1,000 nodes, and compare the average A.R. and Time-to-Solution (T.t.S.) of each case in 10 independent trials. We use 12-dimensional sub-QUBO and the number of sub-QUBOs is 50\% of the original QUBO size. It can be seen from Fig.~\ref{fig:mxc} (a), for Max-Cut problems, DQAOA with PFS exhibits significantly higher A.R., and BFS also tends to outperform random decomposition. Interestingly, the IFD decomposition yields a lower A.R. For sparse QUBO problems, this may originate from insufficient sub-QUBO sampling driven by impact factors without considering pair interactions. The T.t.S. is the time needed for convergence, which is denoted by the change of the A.R.. It remains within $\pm0.1$ for 5 consecutive cycles. As depicted in Fig.~\ref{fig:mxc} (b), IFD and derived BFS/PFS decomposition can lead the DQAOA to faster convergence.

\subsection{Performance analysis of DQAOA}

\subsubsection{Scaling test on computational resources}
\begin{figure}[!h]
\centering
\includegraphics[width=0.7\linewidth]{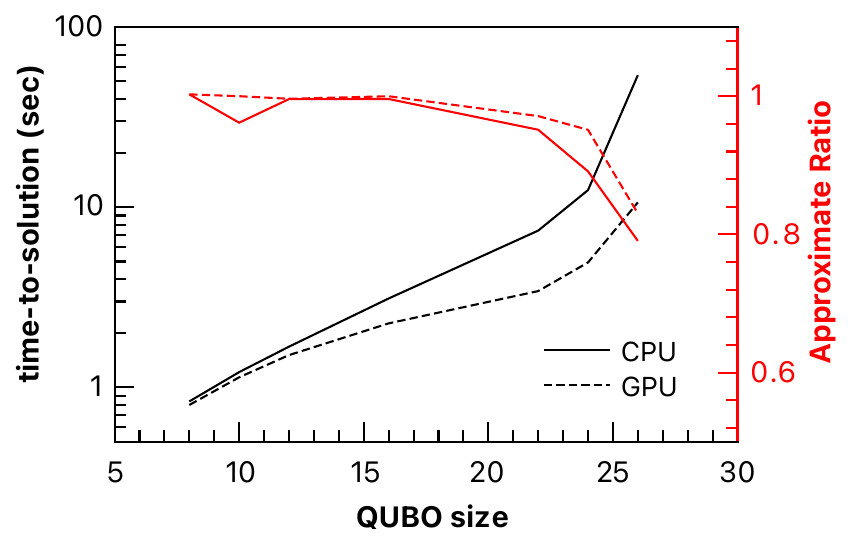}
\vspace*{-\baselineskip}
\caption{\label{fig:single}Performance of Qiskit Aer simulator on a single CPU and a single GPU processor. The time-to-solution is on a log-scale.}
\end{figure}

Since high-performance QPUs have not yet been widely applied and
integrated into current HPC systems, large-scale quantum simulations
primarily rely on quantum simulators running on CPUs or
GPUs. Currently, most of the quantum simulators, such as qiskit-aer
and nwq-sim, support MPI-based parallel execution on CPUs and have
made significant advancements in GPU
acceleration. Figure~\ref{fig:single} shows the performance comparison
of Qiskit Aer simulators when executing QAOA on a single CPU and GPU
processor to solve QUBO problems from 8 to 26 dimensions. T.t.S. in the figure (and following figures without further statement) is shown on a log-scale. The computational advantage of a GPU becomes apparent when the problem size
is greater than 16.

\begin{figure*}[!h]
\centering
\includegraphics[width=0.9\linewidth]{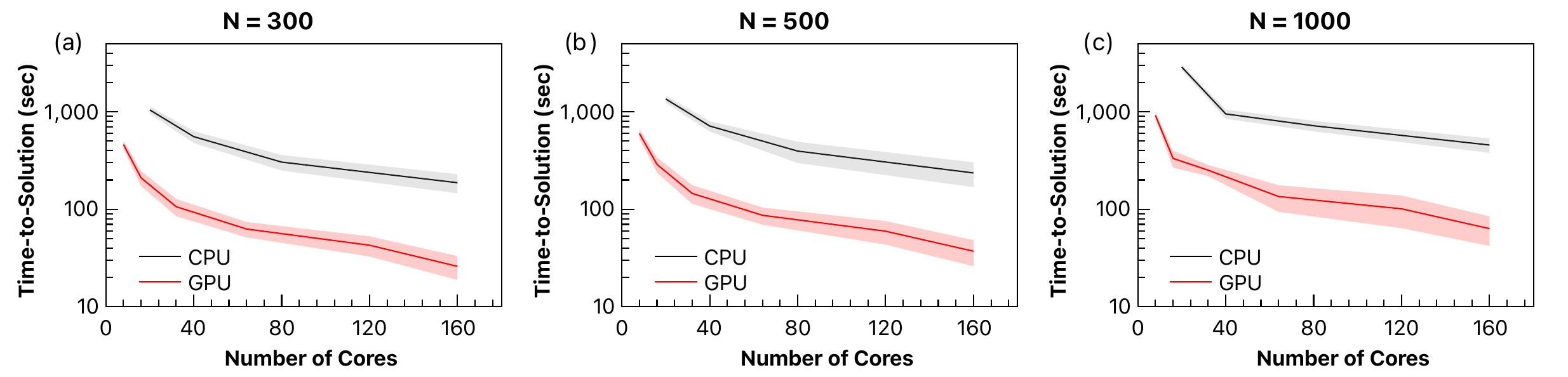}
\vspace*{-\baselineskip}
\caption{\label{fig:scaling-ii}Scaling test of DQAOA on up tp 160 CPU cores/GPUs. (a) to (c) represent QUBO problem 300, 500 and 1000,
  respectively. All the scaling tests run with 10 independent trials
  to get average time-to-solution and error bars. The time-to-solution is on a log-scale.}
\end{figure*}

\begin{figure*}[!h]
\centering
\includegraphics[width=0.9\linewidth]{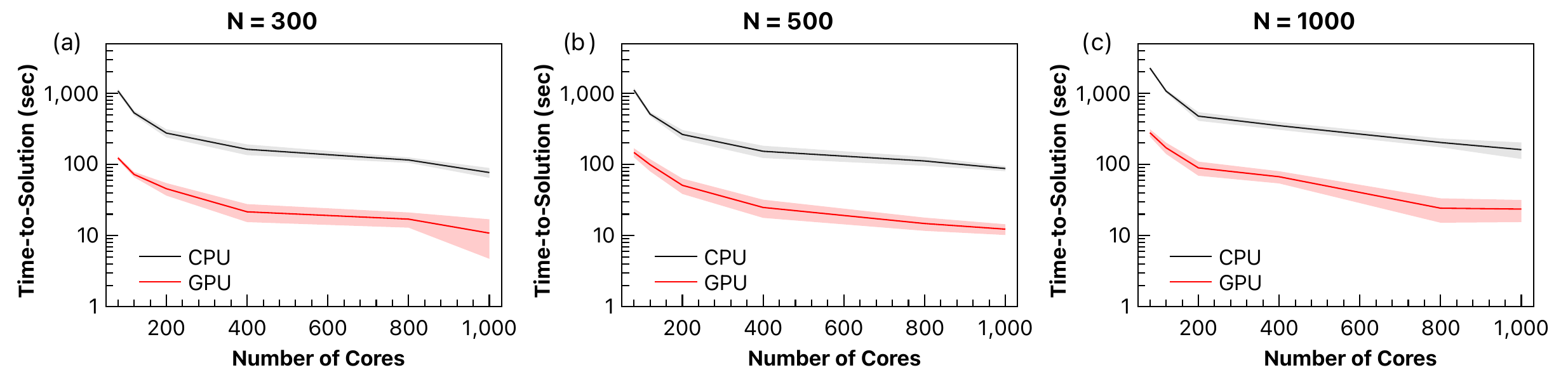}
\vspace*{-\baselineskip}
\caption{\label{fig:scaling-ii-1000}Scaling test of DQAOA on up to 1000 CPU cores/GPUs. (a) to (c) represent QUBO problem 300, 500 and 1000,
  respectively. All the scaling tests run with 10 independent trials
  to get average time-to-solution and error bars. The time-to-solution is on a log-scale.}
\end{figure*}

Figure~\ref{fig:scaling-ii} shows the DQAOA performance on CPUs and GPUs when solving QUBO problems with 300, 500 and 1,000 dimensions. All the scaling tests are on 64-core AMD EPYC CPUs and AMD Instinct MI250X GPUs on the Frontier HPC system. Here, we used 160 sub-QUBOs of size 20 to decompose all of the original QUBO problem. The results are obtained from 10 independent trials with the same initializations. With the same hyperparameters, DQAOA running on GPUs consistently achieves faster convergence than on CPUs, primarily due to the accelerated quantum circuit simulations enabled by GPU parallelism. As the number of available devices increases, up to 160 in this case (equals to the number of sub-QUBOs), the T.t.S. continues to decrease, indicating strong scalability. While CPUs also exhibit performance improvements with additional resources, their scaling efficiency is limited by higher memory access latency and lower parallel processing capabilities. The results suggest that GPUs provide a substantial advantage for solving large-scale QUBO problems with DQAOA. When using more sub-QUBOs (1000 sub-QUBOS), more CPU cores and GPUs are required to address the challenge of scaling up. As shown in Figure~\ref{fig:scaling-ii-1000}, both CPU and GPU show excellent scalability. By comparing the results in Fig.~\ref{fig:scaling-ii} and Fig.~\ref{fig:scaling-ii-1000}, using more sub-QUBOs can significantly accelerate convergence. The availability of large-scale parallel computing resources is expected to significantly enhance the capability of the DQAOA algorithm in solving extremely large QUBO problems.

\subsubsection{Scaling test on different DQAOA hyperparameters}
% \begin{figure*}[h!]
%     \centering
%     \includegraphics[width=0.8\linewidth]{Figures/DQAOA-QFw.pdf}
%     \caption{The workflow of DQAOA integrating with QFw.}
%     \label{fig:dqaoa-qfw}
% \end{figure*}
When using DQAOA to solve large QUBO problems, the selection of hyperparameters, particularly the number and size of sub-QUBOs, plays a crucial role in optimizing performance. If the sub-QUBOs are too small, a large number of them may be required to fully cover the original problem. However, since the number of sub-QUBOs that can be processed in each cycle is typically constrained by available computing resources, increasing the size of sub-QUBOs becomes necessary to reduce their overall count. This, in turn, extends the runtime of each individual QAOA solver. Striking the right balance between sub-QUBO size and number is essential, as it can also minimize the number of cycles needed for DQAOA to converge, which is a key determinant of overall efficiency.

\begin{figure*}[!h]
    \centering
    \includegraphics[width=1.0\linewidth]{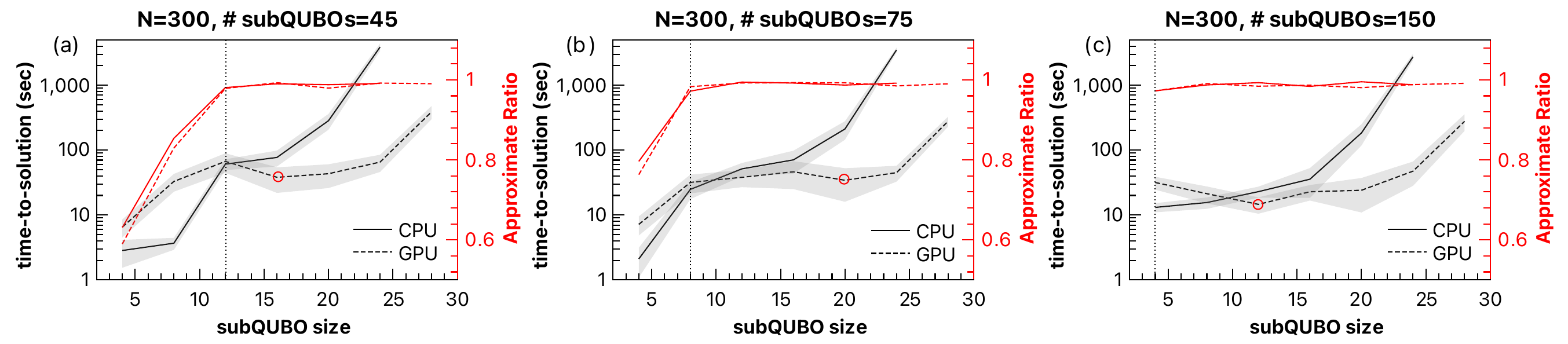}
    \caption{\label{fig:n300}DQAOA simulations for $N=300$. The number
      of sub-QUBOs are (a) 45 (b) 75 and (c) 150. The time-to-solution is on a log-scale.}
\end{figure*}

\begin{figure*}[!h]
    \centering
    \includegraphics[width=1.0\linewidth]{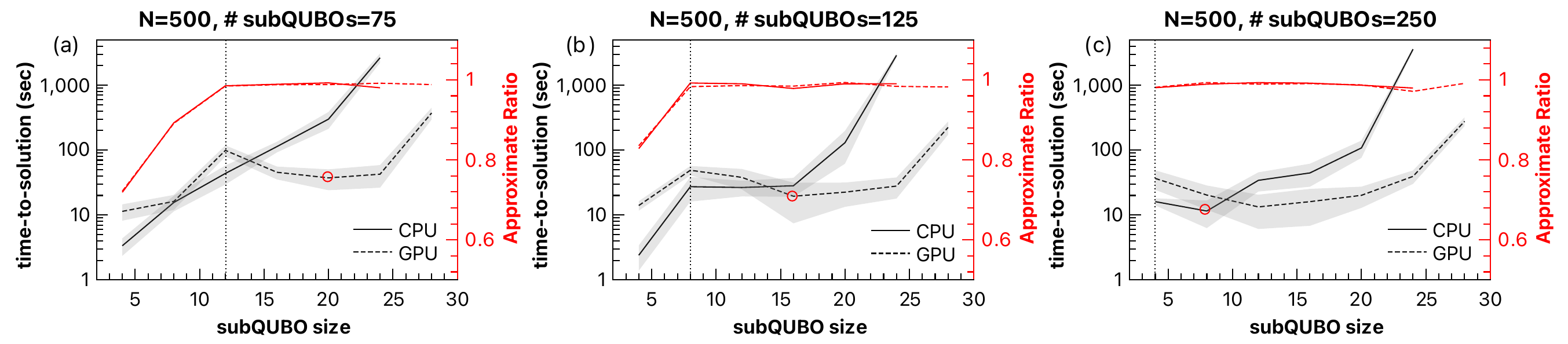}
    \caption{\label{fig:n500}DQAOA simulations for $N=500$. The number
      of sub-QUBOs are (a) 75 (b) 125 and (c) 250. The time-to-solution is on a log-scale.}
\end{figure*}

\begin{figure*}[!h]
    \centering
    \includegraphics[width=1.0\linewidth]{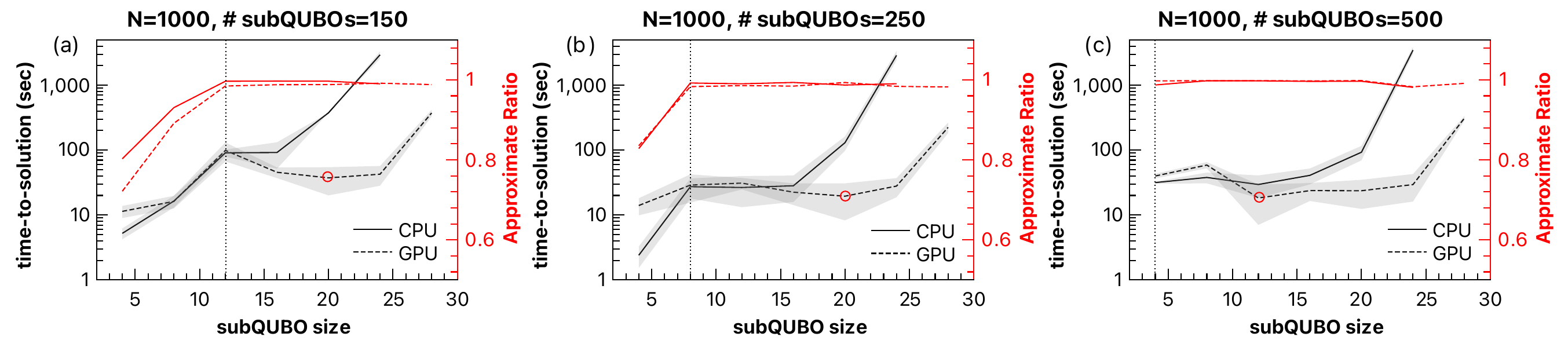}
    \caption{\label{fig:n1000}DQAOA simulations for $N=1000$. The
      number of sub-QUBOs are (a) 150 (b) 250 and (c) 500. The time-to-solution is on a log-scale.}
\end{figure*}

Figures~\ref{fig:n300} to ~\ref{fig:n1000} show the scaling test on
different sub-QUBO sizes and numbers of sub-QUBOs for QUBOs (size:
300, 500 and 1000).  The number of sub-QUBOs is set to 15\%, 25\% and
50\% of the original QUBO size, and the sub-QUBO size ranges from 4 to
28. To prevent computational resource bottlenecks, we reserved 64
nodes on the Frontier HPC system, which provides up to 4096 CPUs and
512 GPUs for the scaling tests, and the number of CPU/GPU used for test is equals to the number of subQUBOs.

As shown in Figures~\ref{fig:n300}, ~\ref{fig:n500}, and
~\ref{fig:n1000}, when using fewer sub-QUBOs, DQAOA requires larger
sub-QUBOs to converge effectively. Conversely, a sufficient number of
sub-QUBOs reduces the dependence on larger sub-QUBO sizes. Overall,
larger sub-QUBOs provide better coverage of the original problem,
thereby reducing the number of cycles required for DQAOA to
converge. However, this increases the computational cost of solving
each QAOA instance. The advantage of GPUs becomes more pronounced as
the problem size increases. DQAOA performance on CPUs decays
monotonically as the sub-QUBO size increases, while the performance on
GPUs has convex formulation. As indicated by the red circles in the
figures, the optimal performance of DQAOA is typically achieved at a
balance between sub-QUBO size and number on GPUs. However, when the
number of GPUs is insufficient to solve all sub-QUBOs simultaneously,
some QAOA instances are queued, leading to suboptimal performance with
extended time-to-solution. In such cases, when CPUs are more readily
available, using smaller sub-QUBOs with a greater number of sub-QUBOs
can serve as an alternative strategy to achieve high efficiency.

\subsubsection{Performance on large-scale QUBO problems}
\begin{figure}[!h]
    \centering
    \includegraphics[width=0.6\linewidth]{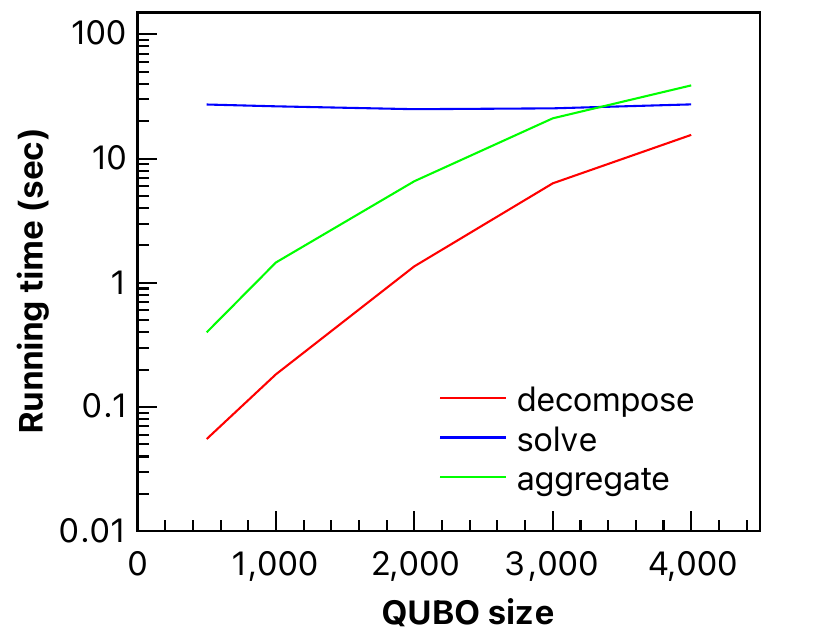}
    \caption{Average decomposing, solving and aggregating time in each
      DQAOA cycle for large QUBO problems. The running time is on a log-scale.}
    \label{fig:large}
\end{figure}

In the discussion above, we leverage GPU to accelerate the quantum
simulation for QAOA instances. However, as the problem size increases,
the efficiency of the decomposition and aggregation policy used in
DQAOA tends to degrade. As previously noted, the computational
complexity of the IFD decomposition and aggregation processes is
$O(N^3+mk^2)$ and $O(mkN^2)$, respectively. Figure~\ref{fig:large}
illustrates how the average runtime of the decomposition, sub-QUBO
solving, and aggregation steps in each DQAOA cycle evolves with
increasing problem size $N$. Here, we fix both the size and number of
sub-QUBOs to 20 and 500, respectively, and conduct all the numerical
experiments using 500 CPU cores. The aggregation becomes the most
computationally expensive step and the runtime of decomposition also
significantly increases as the problem size increases. This suggests
that both the number and size of sub-QUBOs must be carefully managed
to maintain the scalability of the DQAOA algorithm.

\begin{figure*}[h!]
    \centering
    \includegraphics[width=0.8\linewidth]{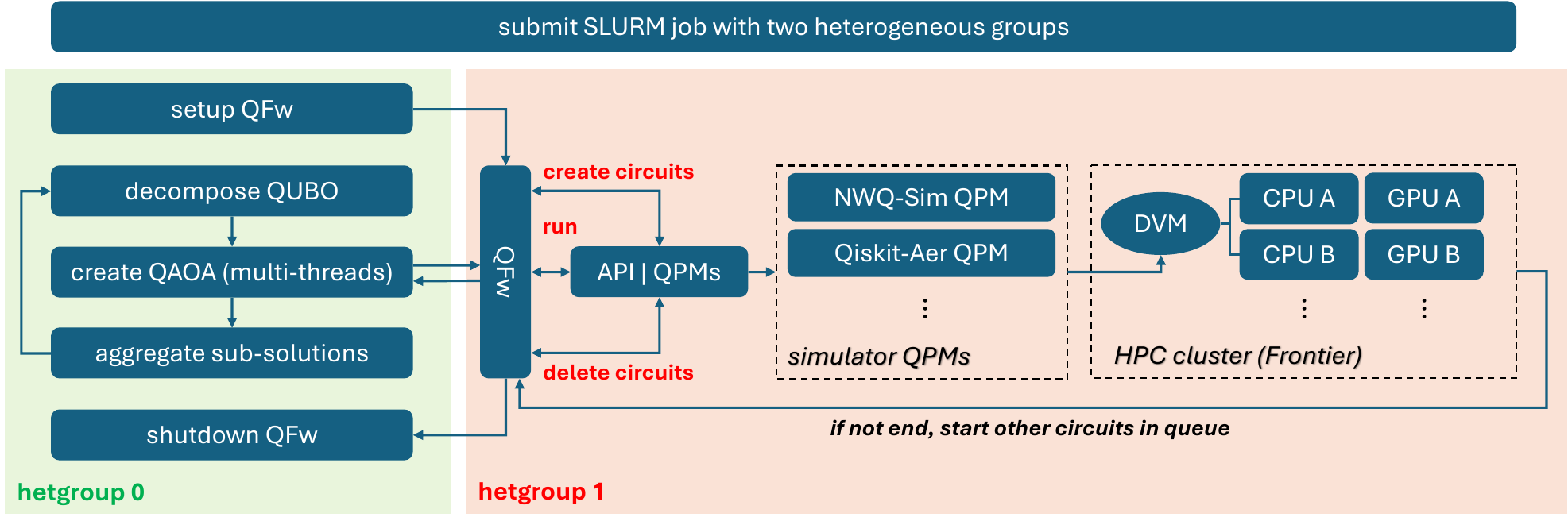}
    \vspace*{-\baselineskip}
    \caption{The workflow of DQAOA integrating with QFw.}
    \label{fig:dqaoa-qfw}
\end{figure*}

\subsection{Integration of DQAOA and QFw}

As the problem size increases, achieving high-quality solutions
requires a greater number of and larger sub-QUBOs. The parallelization
strategy shown in Figure~\ref{fig:dqaoa-mpi} distributes individual
sub-problems across processors, which tends to result in long time
blocking on processors when the number of sub-problems grows rapidly
and the computational cost of solving each sub-problem becomes
excessively high. With the development of HPC-QC systems, DQAOA is
expected to undergo a paradigm shift. Instead of the traditional
MPI-based distribution of sub-problems to individual processors,
QFw introduces a more flexible and scalable approach. In this framework,
quantum ansatze generated by QAOA are asynchronously submitted to one
or multiple processors for simulation. The results are then returned
to a task manager, which updates parameters and generates new ansatze
for subsequent executions. This fine-grained parallelism enhances
resource utilization and management, overcoming the limitations of
sub-problem-based parallelization.

Figure~\ref{fig:dqaoa-qfw} illustrates how DQAOA integrates with
QFw. QFw is designed to leverage the full power of HPC resources while
supporting quantum simulations across multiple
backends~\cite{shehata2024frameworkintegratingquantumsimulation,RN246}.
When submitting a DQAOA job with two heterogeneous groups (hetgroup 0
and hetgroup 1), all the classical applications/operations are
executed on hetgroup 0, while QFw is deployed on the other (hetgroup
1) to manage quantum circuit executions. In this framework, the
original QUBO problem is decomposed and each sub-QUBO, together with
its run options, is wrapped into a single QAOA instance. All the
quantum executions in QAOA are passed to QFw on hetgroup 1 and
invoke the QPM-API to execute with the selected backend (simulator
QPMs, e.g., Qiskit Aer~\cite{RN250}, NWQ-sim~\cite{RN249}, etc.). QFw
distributes these executions via MPI across available computational
resources (CPUs, GPUs, etc.) across multiple nodes. Once the execution
completes, QFw collects the results (sub-solutions) and sends them
back to hetgroup 0 for parameter updating to generate new quantum
circuits. After gathering all the sub-solutions, DQAOA updates the
original solution and starts a subsequent cycle.

%\textcolor{blue}{Srikar, please add your results with QFw}

\begin{figure}
    \centering
    \includegraphics[width=0.9\linewidth]{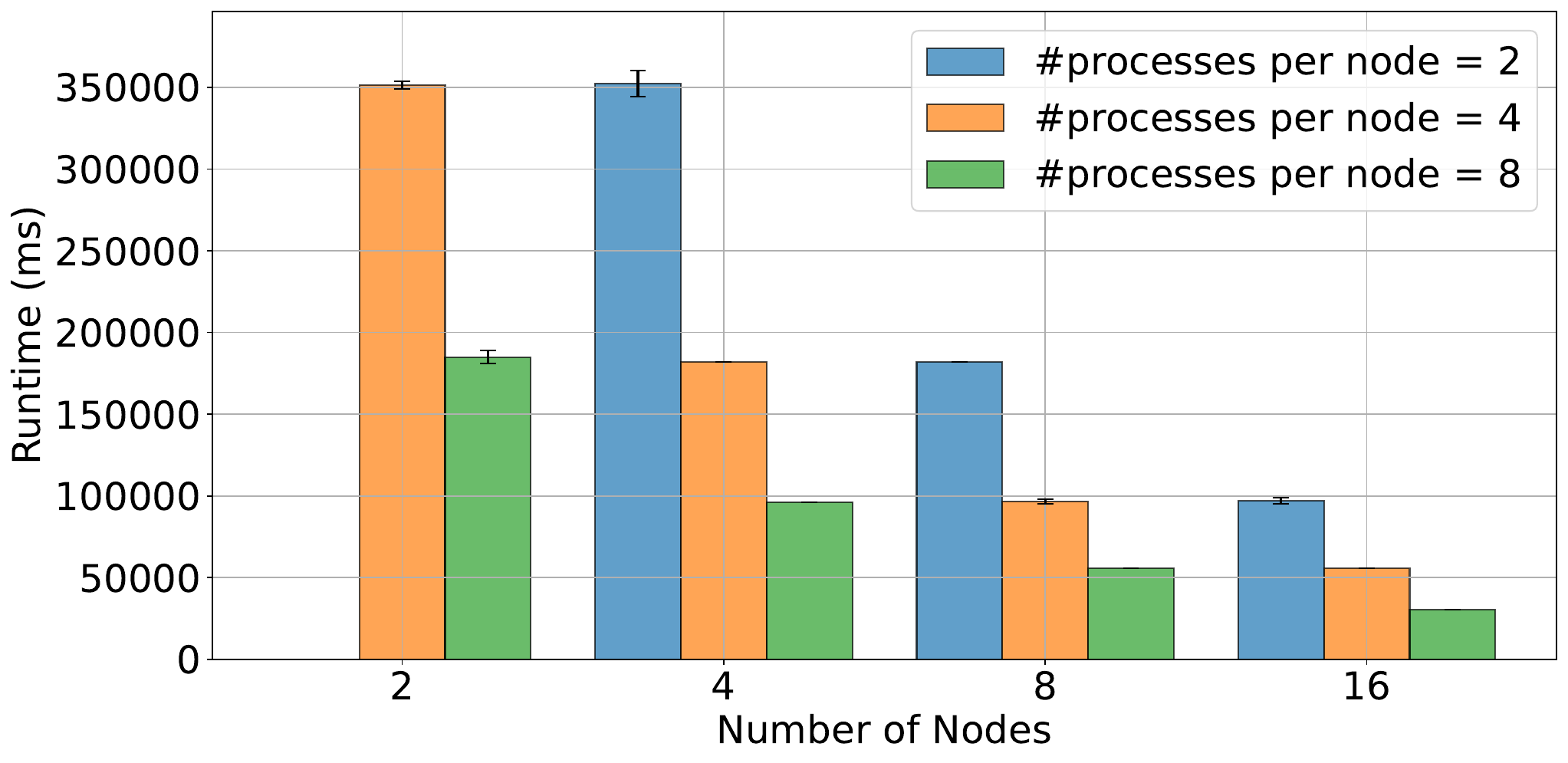}
    \caption[Multi-node GHZ simulations with QFw]{Simulation of the
      30-qubit GHZ circuit on multiple CPUs using QFw. While
      Qiskit-Aer does not natively support MPI-based execution, QFw
      enables scalable simulations with backends like NWQ-Sim that
      support MPI, requiring minimal additional programming effort.}
    \label{fig:multi-ghz-qfw}
\end{figure}
\begin{figure}
    \centering
    \includegraphics[width=0.9\linewidth]{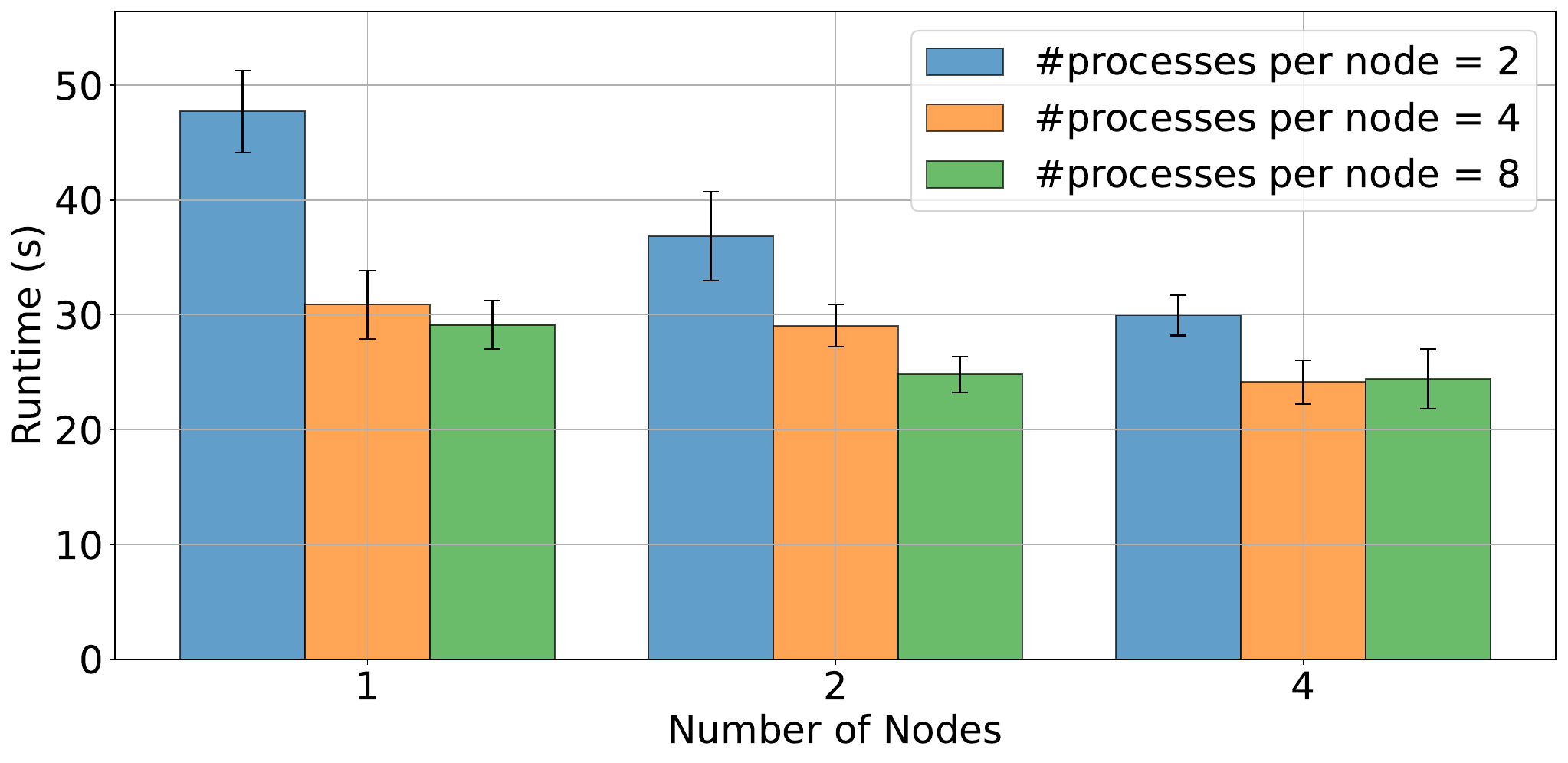}
    \caption[Multi-node QAOA simulations with QFw]{Simulation of the
      QUBO-20 QAOA circuit on multiple CPUs using QFw. This serves
      as an example of how QFw enables scalable execution of
      variational algorithms using MPI, with minimal programming
      overhead --- similar to the abstraction demonstrated for GHZ in
      Fig.~\ref{fig:multi-ghz-qfw}.}
    \label{fig:multi-qaoa-qfw}
\end{figure}

Figure~\ref{fig:multi-ghz-qfw} demonstrates strong scaling for GHZ-30 on Frontier, with runtime dropping from \(352\,\text{s}\) (2 nodes, 4 procs/node) to \(30\,\text{s}\) (16 nodes, 8 procs/node)—an \(11.7\times\) speedup (\(91.5\%\) reduction). Figure~\ref{fig:multi-qaoa-qfw} shows similar behavior for QAOA-20, improving from \(47.5\,\text{s}\) (1 node, 2 procs/node) to \(24.8\,\text{s}\) (4 nodes, 8 procs/node), a \(1.9\times\) speedup (\(47.8\%\) reduction). These results highlight QFw’s scalability for both single-run and variational workloads.
% %
% \fm{Need to interpret results: Is it scaling? Can you run on more MP processes/nodes?}
% %
% \sri{better interpretation now, but I cannot run higher MPI runs due to personal time constraints.}

\section{\textbf{Conclusion}}

%{\textcolor{blue} All, 1st author}
%In this work, we have ...

In this work, we demonstrated the scalability and efficiency of
GPU-accelerated DQAOA for solving high-dimensional combinatorial
optimization problems on large-scale HPC ecosystems. By integrating
advanced problem decomposition techniques --- specifically, IFD
decomposition --- and leveraging MPI-based multi-node parallelism on the
Frontier supercomputer, we achieved a significant speedup,
approximately 10 times faster compared to CPU-based
implementations. Our results highlight that the performance of DQAOA
is strongly influenced by the choice of decomposition strategy, where
IFD decomposition, which prioritizes the most influential problem
components, outperforms random decomposition by improving convergence
and solution quality. Furthermore, optimized resource utilization,
including efficient workload distribution across GPU nodes and
balanced quantum-classical orchestration, played a critical role in
achieving scalable performance. This study also provides a pathway for
deploying GPU-accelerated quantum optimization algorithms in practical
applications, bringing us closer to realizing quantum advantage for
large-scale optimization problems.

In the future, integrating DQAOA with the QFw will enable the seamless execution of hybrid quantum-HPC applications, supporting broader use cases across scientific and
industrial domains. As quantum hardware advances, we plan to explore
dynamic scheduling techniques for adaptive workload distribution
across quantum and classical resources, optimizing execution
efficiency. Furthermore, extending GPU-accelerated simulations to
incorporate noise-aware and error-mitigated approaches will be crucial
for bridging the gap between simulated and real-world quantum
performance. These advancements will contribute to the ongoing
development of scalable, high-performance quantum-HPC ecosystems for
next-generation computational challenges.

\section*{Acknowledgements}
%Omitted for Double-Blind Review.
This work was supported in part by NSF CISE-2217020, CISE-2316201, OSI-2120757, PHY-1818914, PHY-2325080.
This research used resources of the Oak Ridge Leadership Computing Facility at the Oak Ridge National Laboratory, which is supported by the Office of Science of the U.S. Department of Energy under Contract No. DE-AC05-00OR22725. 
{\it Notice}: This manuscript has been authored by UT-Battelle, LLC, under contract DE-AC05-00OR22725 with the US Department of Energy (DOE). The US government retains and the publisher, by accepting the article for publication, acknowledges that the US government retains a nonexclusive, paid-up, irrevocable, worldwide license to publish or reproduce the published form of this manuscript, or allow others to do so, for US government purposes. DOE will provide public access to these results of federally sponsored research in accordance with the DOE Public Access Plan (https://www.energy.gov/doe-public-access-plan).

\newpage
\bibliography{refs}
\bibliographystyle{IEEEtran}
%\bibliographystyle{plain}
% \bibliographystyle{unsrt}

% \input{sections/13-appendix.tex}
% \newpage

% \includepdf[pages=-]{sections/DIAQ_Artifact_Evaluation_Report.pdf}
%\newpage

%\input{sections/14-rebuttal-sc24}

\end{document}